\documentclass[nofootinbib,twocolumn,preprintnumbers]{revtex4-1}
\pdfoutput=1
\usepackage{amsmath,amsthm,amssymb,multirow,psfrag}
\usepackage{epsfig}
\usepackage{color}
\usepackage{slashed}
\graphicspath{{./Figures/}}

\begin{document}

\def\lsim{\mathrel{\rlap{\lower4pt\hbox{\hskip1pt$\sim$}}
  \raise1pt\hbox{$<$}}}
\def\gsim{\mathrel{\rlap{\lower4pt\hbox{\hskip1pt$\sim$}}
  \raise1pt\hbox{$>$}}}
\newcommand{\vev}[1]{ \left\langle {#1} \right\rangle }
\newcommand{\bra}[1]{ \langle {#1} | }
\newcommand{\ket}[1]{ | {#1} \rangle }
\newcommand{\ev}{ {\rm eV} }
\newcommand{\kev}{{\rm keV}}
\newcommand{\mev}{{\rm MeV}}
\newcommand{\tev}{{\rm TeV}}
\newcommand{\mpl}{$M_{Pl}$}
\newcommand{\mw}{$M_{W}$}
\newcommand{\Ft}{F_{T}}
\newcommand{\Zparity}{\mathbb{Z}_2}
\newcommand{\BLambda}{\boldsymbol{\lambda}}
\newcommand{\met}{\;\not\!\!\!{E}_T}
\newcommand{\beq}{\begin{equation}}
\newcommand{\eeq}{\end{equation}}
\newcommand{\bea}{\begin{eqnarray}}
\newcommand{\eea}{\end{eqnarray}}
\newcommand{\nn}{\nonumber}
\newcommand{\gev}{{\mathrm GeV}}
\newcommand{\hc}{\mathrm{h.c.}}
\newcommand{\eps}{\epsilon}
\newcommand{\bwt}{\begin{widetext}}
\newcommand{\ewt}{\end{widetext}}
\newcommand{\draftnote}[1]{{\bf\color{blue} #1}}

\newcommand{\cO}{{\cal O}}
\newcommand{\cL}{{\cal L}}
\newcommand{\cM}{{\cal M}}

\newcommand{\fref}[1]{Fig.~\ref{fig:#1}} 
\newcommand{\eref}[1]{Eq.~\eqref{eq:#1}} 
\newcommand{\aref}[1]{Appendix~\ref{app:#1}}
\newcommand{\sref}[1]{Section~\ref{sec:#1}}
\newcommand{\tref}[1]{Table~\ref{tab:#1}}

\title{\LARGE{{\bf The Platinum Channel: \\ Higgs Decays to as many as 8 Leptons}}}
\author{{\bf {Eder Izaguirre$\,^{a}$ and Daniel Stolarski$\,^{b}$}}}

\affiliation{
$^a$Brookhaven National Laboratory, Upton, NY 11973, USA\\
$^b$Ottawa-Carleton  Institute  for  Physics,  Carleton  University,\\
1125  Colonel  By  Drive,  Ottawa,  Ontario  K1S  5B6,  Canada
}

\email{
izaguirre.eder@gmail.com \\
stolar@physics.carleton.ca\\}

\begin{abstract}
We propose a search for Higgs decays with as many as eight leptons in the final state. This signal can arise in a simple model with a hidden vector ($A_d$) that gets mass via a hidden scalar ($h_d$) vacuum expectation value. The 125 GeV Higgs can then decay $H\rightarrow h_d h_d \rightarrow 4A_d\rightarrow 8f$, where $f$ are Standard Model fermions. We recast current searches and show that a branching ratio of $H\rightarrow h_dh_d$ as large as 10\% is allowed. We also describe a dedicated search that could place bounds on BR($H\rightarrow h_dh_d$) as low as $10^{-5}$ using only 36 fb$^{-1}$ of data, with significant improvements coming from greater integrated luminosity. 
\end{abstract}

\maketitle

\section{Introduction} 
\label{sec:intro} 

The discovery of the Higgs boson~\cite{Aad:2012tfa,Chatrchyan:2012xdj} completes the Standard Model (SM), but it also opens up a new avenue to look for deviations from the SM. 
As yet, all measurements of the Higgs have been consistent with the SM~\cite{Khachatryan:2016vau}, but deviations due to beyond the SM physics could have been missed so far if these are at a level below current theoretical and/or experimental uncertainties, or if they manifest in unconventional final states. 
In this paper we present an as yet unattempted measurement that could be done to probe physics beyond the SM. 

The Higgs square operator, $H^\dagger H$, is the only gauge invariant scalar operator of dimension lower than four in the SM. Therefore, it is natural to expect that if there is another sector that talks to the SM, its scalars could couple to the SM via this ``Higgs portal'' operator~\cite{Schabinger:2005ei}. In this work, we posit a very simple hidden sector: a new $U(1)$ gauge boson which acquires mass via a hidden sector Higgs mechanism, and the hidden Higgs has a renormalizable coupling to the SM via the Higgs portal. The new gauge boson generically couples to the SM through the ``vector portal''~\cite{Holdom:1985ag}, $B_{\mu\nu}F_d^{\mu\nu}$, where $B_{\mu\nu}$ is the field strength tensor for SM hypercharge, and $F_d^{\mu\nu}$ is the field strength for the hidden gauge group. 
The phenomenology of a hidden abelian gauge group was first studied in~\cite{Babu:1997st}. 

The model with Higgs and vector portal couplings was studied in the ultra-light regime in~\cite{Ahlers:2008qc,Ng:2014iqa}. It was studied for general Higgs phenomenology in~\cite{Weihs:2011wp}, and it has been most thoroughly studied in the context of Higgs decays to four leptons~\cite{Gopalakrishna:2008dv,Davoudiasl:2013aya,Curtin:2013fra,Falkowski:2014ffa,Curtin:2014cca,Bakhet:2015pqa}. With this model, however, there is a large region of parameter space where decays to more than four leptons are possible. If we take the hidden scalar to be lighter than half the Higgs mass, and the hidden photon to be lighter than half the hidden scalar mass,\footnote{If the hidden photon is heavier than half the hidden scalar mass, decays via off-shell hidden vectors are allowed but will be suppressed by the vector portal coupling and may be smaller than the decay to two SM fermions via mixing with the SM-like Higgs.} 
then the SM Higgs could decay via 
\begin{equation}
H \rightarrow h_d h_d \rightarrow A_d A_d A_d A_d \rightarrow 8 f \, ,
\label{eq:decay}
\end{equation}
where $H$ is the SM Higgs at 125 GeV, $h_d$ and $A_d$ are the hidden sector scalar and vector respectively, and $f$ are SM fermions. The first decay occurs through the Higgs portal operator and current limits allow its branching ratio to be as large as $\mathcal{O}(10\%)$. The second decay is the dominant decay of the hidden sector Higgs if kinematically allowed because of the minimality of the hidden sector. If there were other hidden sector fields then this branching ratio could be reduced, but it is naturally large as long as the hidden gauge coupling is reasonably large. 

The decay of the hidden photon goes via the vector portal coupling even if it is extremely small. The Higgs portal coupling does not mediate hidden vector decays at tree level. If the hidden vector is parametrically lighter than the $Z$, then it dominantly couples to the electromagnetic current, thus giving each hidden photon a significant branching ratio to SM leptons. This branching ratio can be extracted from the $R$ ratio of $e^+e^-$ scattering to hadrons relative to that to muons~\cite{Curtin:2014cca}. This can in turn be extracted from data at low masses~\cite{Curtin:2014cca}, and from three-loop QCD calculation of $R$ at higher masses~\cite{Chetyrkin:2000zk}. 
We call the decay in \eref{decay} the platinum channel because of how spectacular it would be at the LHC. 

Higgs decays to lepton jets~\cite{Falkowski:2010gv} can also arise from this model~\cite{Chang:2013lfa} (see also~\cite{Chang:2016lfq} for Higgs decays to lepton jets in a different model), and the work of~\cite{Chang:2013lfa} studies Higgs decays to leptons where the mass of the $A_d$ is $\sim 1$ GeV so that the final state lepton pairs are very collimated and may be treated as a single detector object. In this work we consider the general case as long as the decays in \eref{decay} are kinematically allowed and explore the phenomenology of this scenario. We find that current constraints on this process are dominated by the CMS multi-lepton searches~\cite{Sirunyan:2017lae} and are quite weak. We also show that there are searches that are very low background and could be performed with current and future data which would explore significant regions of parameter space. 
We therefore hope that this work will spur future study by our experimental colleagues. 

The rest of this paper proceeds as follows. In Section~\ref{sec:model}, we present the details of a simple model that gives rise to this decay. In Section~\ref{sec:constraints} we explore current constraints on the model including a recast of the CMS multilepton search from~\cite{Sirunyan:2017lae}, and in Section~\ref{sec:projections} we show how a dedicated search could significantly improve the limits. 
In Section~\ref{sec:nonmin} we briefly explore non-minimal models that give rise to this scenario, and conclusions are given in Section~\ref{sec:conclusion}.

\section{A Simple Model}
\label{sec:model}

We consider the following hidden sector Lagrangian added to the SM
\begin{equation}
{\cal L}_{\rm hidden}= -\frac{1}{4}F^{\mu\nu}F_{\mu\nu} +|D_\mu h_d|^2 -V(h_d^\dagger h_d) \, ,
\label{eq:lhid}
\end{equation}
where $h_d$ is the hidden (or dark) sector Higgs, and $F_{\mu\nu}$ is the field strength tensor for the hidden $U(1)$ gauge boson $A_d$. The $h_d$ has unit charge under the hidden $U(1)$. $V(h_d^\dagger h_d)$ is the usual wine bottle potential with negative mass squared term so that $h_d$ gets a vacuum expectation value (vev) even in the absence of portal operators. We also add a portal Lagrangian:
\begin{equation}
{\cal L}_{\rm portal} =  \frac{\epsilon}{2\cos\theta_w}F^{\mu\nu}B_{\mu\nu} + \lambda h_d^\dagger h_d H^\dagger H \, ,
\end{equation}
where $H$ is the SM Higgs and $B_{\mu\nu}$ is the field strength for SM hypercharge. Current limits on this model require both $\lambda$ and $\epsilon$ to be small as we will see in detail below, so we work to first order in both. Detailed formulae for the mixings and couplings in this model can be found, for example, in~\cite{Gopalakrishna:2008dv,Chang:2013lfa,Curtin:2014cca}. Here we state the results for the processes of interest in our study.

Both the SM Higgs and the hidden Higgs get vevs in the absence of the portal coupling:
\begin{equation}
\langle h_d \rangle \approx \frac{v_d}{\sqrt{2}}\;\;\; \;\;\; 
\langle H \rangle \approx 
\frac{1}{\sqrt{2}}\left(
\begin{matrix}
    0      \\
     v
\end{matrix}
\right)
\end{equation}
with $v\approx 246$ GeV. The Higgs portal coupling shifts the vevs by $\mathcal{O}(\lambda)$, and it induces mixing between the SM and hidden Higgses, which in turn allows the SM Higgs to decay to two hidden vectors. If kinematically allowed, the tree level width for this decay is given by:
\begin{eqnarray}
\Gamma(H \rightarrow A_d A_d) = \frac{\lambda^2}{32\pi}\frac{v^2}{m_H}\left(1-\frac{m_{h_d}^2}{m_H^2}\right)^{-2}\nonumber\\
\times \sqrt{1-\frac{4m_{A_d}^2}{m_H^2}} \left(1-\frac{4m_{A_d}^2}{m_H^2}+\frac{12m_{A_d}^4}{m_H^4} \right)
\, .
\label{eq:Haa}
\end{eqnarray}
The decay of the Higgs to two hidden Higgses is mediated by the Higgs portal coupling with a Higgs vev insertion
\begin{equation}
\Gamma(H \rightarrow h_d h_d) =\frac{\lambda^2 v^2}{32\pi m_H}\sqrt{1-\frac{4m_{h_d}^2}{m_H^2}}\, .
\label{eq:Hhh}
\end{equation}
Therefore the branching ratio to hidden scalars is typically comparable to that to hidden vectors.

This model can also give Higgs decay to $Z$ and $A_d$ which would go through the vector portal. Constraints require $\epsilon \lesssim  10^{-3}$ (see below), and this decay is further suppressed by $m_A^2/m_Z^2$, so it is negligible in the parameter space of interest. The $Z$ can also decay as $Z\rightarrow A_d h_d$ which was studied in detail in~\cite{Blinov:2017dtk}. While the LHC is not presently sensitive to this decay in this model, it may become sensitive in the future. 

The branching ratio of the SM-like Higgs decay to hidden scalars is given by:
\begin{equation}
{\rm BR}(H\rightarrow h_dh_d) \approx \frac{\Gamma(H \rightarrow h_d h_d) }{\Gamma_H^{\rm SM}} \approx 0.1\% \left(\frac{\lambda}{10^{-3}}\right)^2 \, ,
\label{eq:BRHhh}
\end{equation}
where the first approximation is that the hidden sector does not significantly contribute to the total width, and the second is assuming that the hidden scalar mass is well below half the Higgs mass. 

With this minimal hidden sector, the only decay of the hidden Higgs that is not suppressed by small couplings is that to two hidden vectors as long as it is kinematically allowed. So in that regime
\begin{equation}
{\rm BR}(h_d \rightarrow A_d A_d) \approx 100\%, \;\;\; \;\;\; m_{h_d} > 2 m_{A_d}\, .
\label{eq:BRHaa}
\end{equation}
One could expand the hidden sector to include, for example, a dark matter candidate~\cite{Pospelov:2007mp}. This could change some of the phenomenology, but we leave more complicated models to future work.

The hidden vector couples to the electromagnetic current with strength $\epsilon e$ and thus couples democratically to electromagnetic charge. It also couples to the $Z$ current, but that is suppressed by $m_{A_d}^2/m_Z^2$ which is small in the region of parameter space we are interested in. The branching ratio of the hidden vector to leptons ($e$ and $\mu$) was calculated very precisely in~\cite{Curtin:2014cca} 
and is typically large as long as $m_{A_d}$ is not near a QCD resonance. 
In this preliminary collider study we use tree-level branching ratios keeping in mind that this will not be a suitable approximation near QCD resonances. 

From the computations in~\cite{Curtin:2014cca}, we can also compute the lifetime of the $A_d$ very precisely, but in the range we are interested, it is approximately given by 
\begin{equation}
\Gamma \simeq \frac{\epsilon^2 \,m_{A_d}}{8\pi}\, ,
\end{equation}
which translates to a lifetime of 
\begin{equation}
c\tau \simeq 5\times 10^{-8}\,{\rm m}\left(\frac{10^{-4}}{\epsilon}\right)^2 \left(\frac{10\,{\rm GeV}}{m_{A_d}}\right)\, ,
\end{equation}
so the $A_d$ decays promptly as long as $\epsilon \gtrsim 10^{-6}$, which is the range we will focus on here. If the hidden photon has a macroscopic lifetime, then the current constraints as well as experimental challenges for finding it are quite different, and we leave the small $\epsilon$ case with displaced decays to future work.

\section{Current Constraints}
\label{sec:constraints}

We first look at constraints on direct production of the hidden sector fields. If the hidden vector is lighter than the hidden scalar, then dark photon constraints can be straightforwardly applied to this scenario. For 1 GeV $\lesssim m_{A_d} \lesssim 10$ GeV, the strongest constraints come from BABAR~\cite{Lees:2014xha} through resonant production of $A_d$ and decay into SM leptons, and set a bound on the kinetic mixing parameter $\epsilon$, namely
\begin{equation}
\epsilon \lesssim {\rm few}\times 10^{-4}, \;\;\; 1 \, {\rm GeV} \lesssim m_{A_d} \lesssim 10\, {\rm GeV} \, .
\end{equation}
Regions close to narrow QCD resonances have much weaker bounds. In this work, we therefore, do not consider $m_{A_d}$ very close to the mass of the $\phi$, $J/\psi$ and $\Upsilon$ resonances. For larger masses, the leading bounds on $\epsilon$ come from LHCb~\cite{Aaij:2017rft} through a dilepton resonance analysis, where the bounds are
\begin{equation}
\epsilon \lesssim 10^{-3}, \;\;\; 10 \, {\rm GeV} \lesssim m_{A_d} \lesssim 40\, {\rm GeV} \, .
\end{equation}
These bounds apply to prompt decays of the hidden vector, the case we consider here, and we see that there are at least two decades of allowed parameter space where the hidden photon is prompt and not excluded.  

In the mass range of interest for the hidden scalar, $10\;\text{GeV} \lesssim m_{h_d} \lesssim 60 \;\text{GeV}$, the strongest limits on direct production of the $h_d$ via its mixing with the SM Higgs come from LEP. The $h_d$ will dominantly decay  to two $A_d$, which then each decay to a pair of SM fermions. Most searches do not look for this particular decay channel, so the bounds are quite weak. The strongest bound comes from the decay mode independent search at OPAL~\cite{Abbiendi:2002qp}, which places a limit on $\sin^2\theta_h$ where $\theta_h$ is the mixing angle between the SM-like and hidden Higgs. This limit varies from $\sim 0.05$ at low mass to $\sim 0.6$ at high mass. In our model, 
\begin{equation}
\sin\theta_h \approx \frac{\lambda v v_d}{m_H^2-m_{h_d}^2}\, .
\end{equation} 
We can write $m_{A_d} = g_d v_d$ and then use this search to set limits on the scalar portal coupling $\lambda$ as a function of $m_{A_d}$, $m_{h_d}$ and $g_d$. The limits are inversely proportional to $g_d$, the hidden gauge coupling, and 
this search only sets limits for very small values of the hidden gauge coupling, 
$g_d \lesssim 10^{-2}$. 
Searches for topologies of the type~\cite{Abbiendi:2002in,Schael:2006cr}
\begin{equation}
e^+e^-\rightarrow H_2 Z \rightarrow H_1 H_1 Z \rightarrow 4\,\text{SM}+Z
\end{equation}
could be sensitive to direct production of $h_d$ if we identify $H_2=h_d$ and $H_1=A_d$. These searches, however, do not put any bounds on the scenario, mainly because they require specific final states, and the branching ratio of the $A_d$ to any particular SM state is somewhat small. 

LHC constraints arising from decays of the 125 GeV Higgs can be set because the mixing of the $h_d$ and $H$ induces decays to $A_dA_d$ which can result in the Higgs decay to four leptons~\cite{Gopalakrishna:2008dv,Davoudiasl:2013aya,Curtin:2013fra,Falkowski:2014ffa,Curtin:2014cca,Bakhet:2015pqa} as shown in \eref{Haa}. This has been searched for at ATLAS~\cite{Aad:2015sva,Aaboud:2018fvk} and CMS~\cite{Khachatryan:2015wka}, with the strongest bounds coming form the recent 13 TeV ATLAS search~\cite{Aaboud:2018fvk}. These limits are shown as the dashed red lines in \fref{Haa}, and are simply the limits shown in Fig.~10 of~\cite{Aaboud:2018fvk}. There are also searches with $\tau$'s and $b$'s in the final state~\cite{Aad:2015oqa,Khachatryan:2017mnf}, but those do not set a non-trivial limit because of significantly larger background than searches with muons or electrons. 

\begin{figure*}[tb]
\centering
\begin{minipage}[c]{\textwidth}
\includegraphics[width=0.7\textwidth]{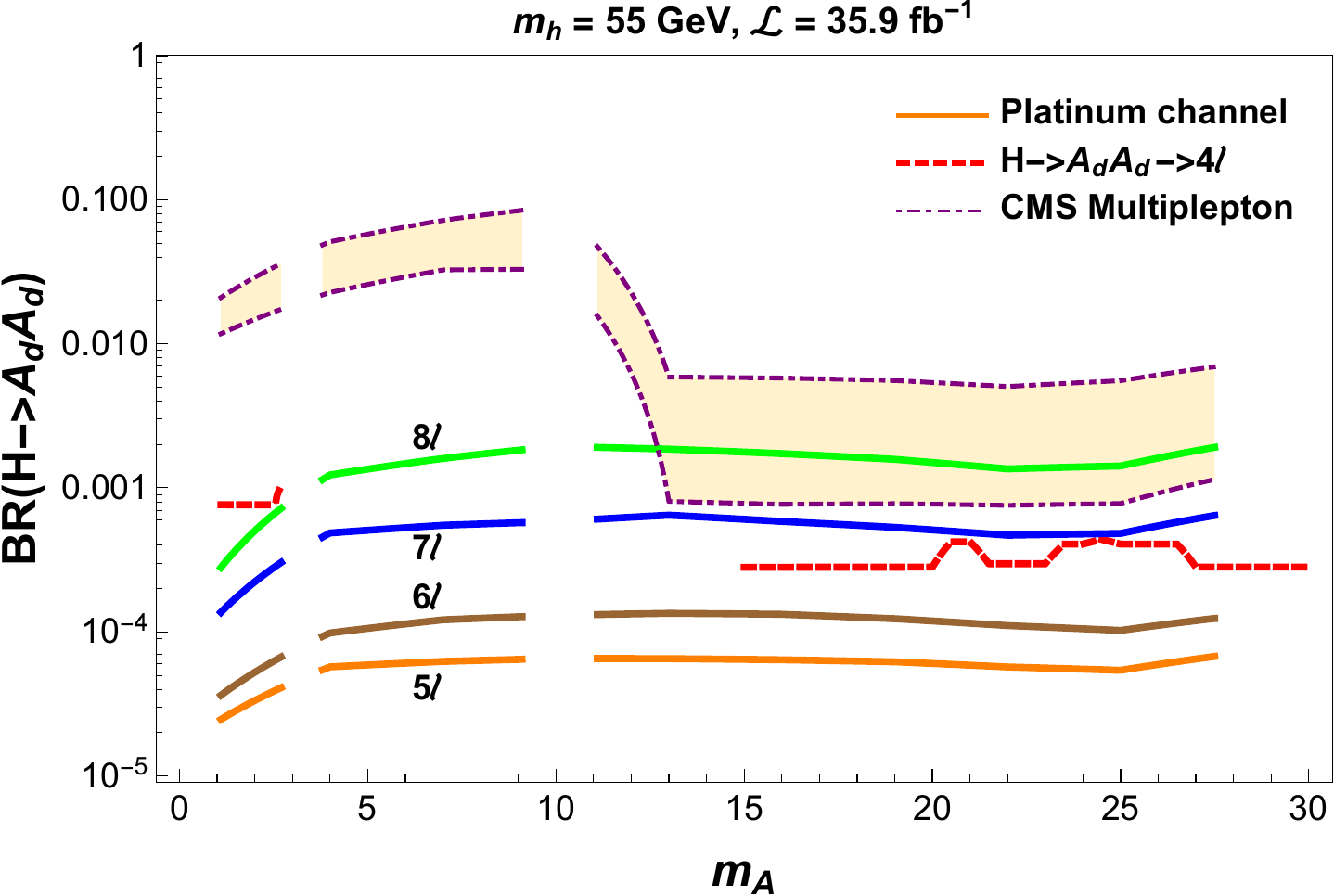}
\end{minipage}
\hfill
\caption{Current and projected limits on the on the hidden sector model considered in this work. The horizontal axis is the hidden vector mass, $m_A$, and the vertical is the branching ratio of the SM-like Higgs to two dark vectors. 
The red dashed curves are limits from the channel $H\rightarrow A_dA_d\rightarrow 4\ell$ from~\cite{Aaboud:2018fvk}. The dot-dashed purple curves are recasted limits on $H\rightarrow h_dh_d\rightarrow 4A_d$ from the CMS multi-lepton search~\cite{Sirunyan:2017lae}, converted to a limit on BR($H\rightarrow A_dA_d$) using Eqs.~\eqref{eq:Haa} and~\eqref{eq:Hhh}. The yellow band parameterizes the uncertainty due to lepton efficiency, see text for details. The solid curves are the \textit{projected} limits from the proposed searches with $\geq$ 5-8 leptons going from bottom to top. Here the mass of the hidden Higgs $h_d$ is set to 55 GeV, but the limits are fairly insensitive to that parameter. The projections use an integrated luminosity of 35.9 fb$^{-1}$. We do not present projections for the hidden photon mass near the $\phi$, $J/\psi$ or $\Upsilon$ resonances.
}
\label{fig:Haa}
\end{figure*}

Finally, we consider the cascade process that can give rise to the decay, $H \rightarrow h_d h_d\rightarrow 4A_d$. This can be constrained by the CMS multilepton study from~\cite{Sirunyan:2017lae}, whose signal regions are potentially applicable to this topology as they require low $p_T$ leptons as well as no missing energy. We recast the limit from~\cite{Sirunyan:2017lae} to set a bound on the model considered here, but we note that because this is a recast, there are significant uncertainties on our limit. We simulate Higgs production at LHC13 using the model from~\cite{Curtin:2014cca} in {\sc MadGraph5\_aMC@NLO}~\cite{Alwall:2014hca}. Higgs production through gluon fusion is simulated at tree-level with an effective gluon-gluon-Higgs vertex, and then the Higgs is forced to decay to $h_d$ pairs, which are then allowed to decay inclusively. We shower and hadronize events using {\sc Pythia8.2}~\cite{Sjostrand:2014zea}. 
While our strategies will focus on leptons, we must shower and hardronize the partons in order to approximate the isolation requirements imposed by experiments. We ignore detector effects in this preliminary study, but we note that these can be important considering the low $p_T$ thresholds we use and the high pile-up environment of the LHC. 

In order to derive the constraints from the CMS search, we must apply
lepton identification efficiencies, which are somewhat small for leptons with
low $p_{\rm T}$. Because~\cite{Sirunyan:2017lae} only provides the
low-$p_{\rm T}$ lepton tagging efficiencies for the most pessimistic working
point, we must use the pessimistic values and obtain a conservative result. The
true signal efficiency is almost certainly better than what we find, because~\cite{Sirunyan:2017lae} states that a looser set of lepton identification criteria are used for searches with four leptons, but does not specifically
state what these efficiencies are. Therefore, we consider efficiencies of 50\% (100\%) to set a conservative (aggressive) limit. 

We find that the Signal Region (SR) H of~\cite{Sirunyan:2017lae}, which
requires 4 leptons and fewer than two opposite-sign, same-flavor (OSSF) lepton
pairs, is most sensitive to the hidden sector topology we study. Using the $\mathrm{CL}_{\rm s}$
method \cite{Read:2002hq}, we estimate a constraint on this scenario at the 95\% confidence level, which is shown as the dot-dashed purple line in \fref{Haa}, with the yellow band showing our uncertainty due to lepton identification efficiencies. All the constraints in \fref{Haa} are shown for $m_{h_d} = 55$ GeV, but the limits are mostly insensitive to the value of this parameter. 
As discussed above, we use the tree-level branching ratios of the $A_d$ to SM fermions, and we mask our plots when $m_{A_d}$ is near the masses of the $\phi$, $J/\psi$, and $\Upsilon$. From \fref{Haa}, we see that the searches for $H\rightarrow A_dA_d$ are more sensitive to this model than the CMS multi-lepton searches, but, as we will show in the next section, a dedicated search could be more sensitive than both.

The CMS multilepton search is sensitive to the process $H\rightarrow h_d h_d\rightarrow 4A_d$, so we also show the constraints placed on BR($H\rightarrow h_dh_d$) as a function of $m_{h_d}$ in \fref{Hhh}. This branching ratio is sensitive to $m_{A_d}$ and the limits vary from 10\% to $10^{-3}$ depending on the $A_d$ mass and on whether we use aggressive or conservative parameterization for lepton efficiency. 

\begin{figure*}[tb]
\centering
\includegraphics[width=0.7\textwidth]{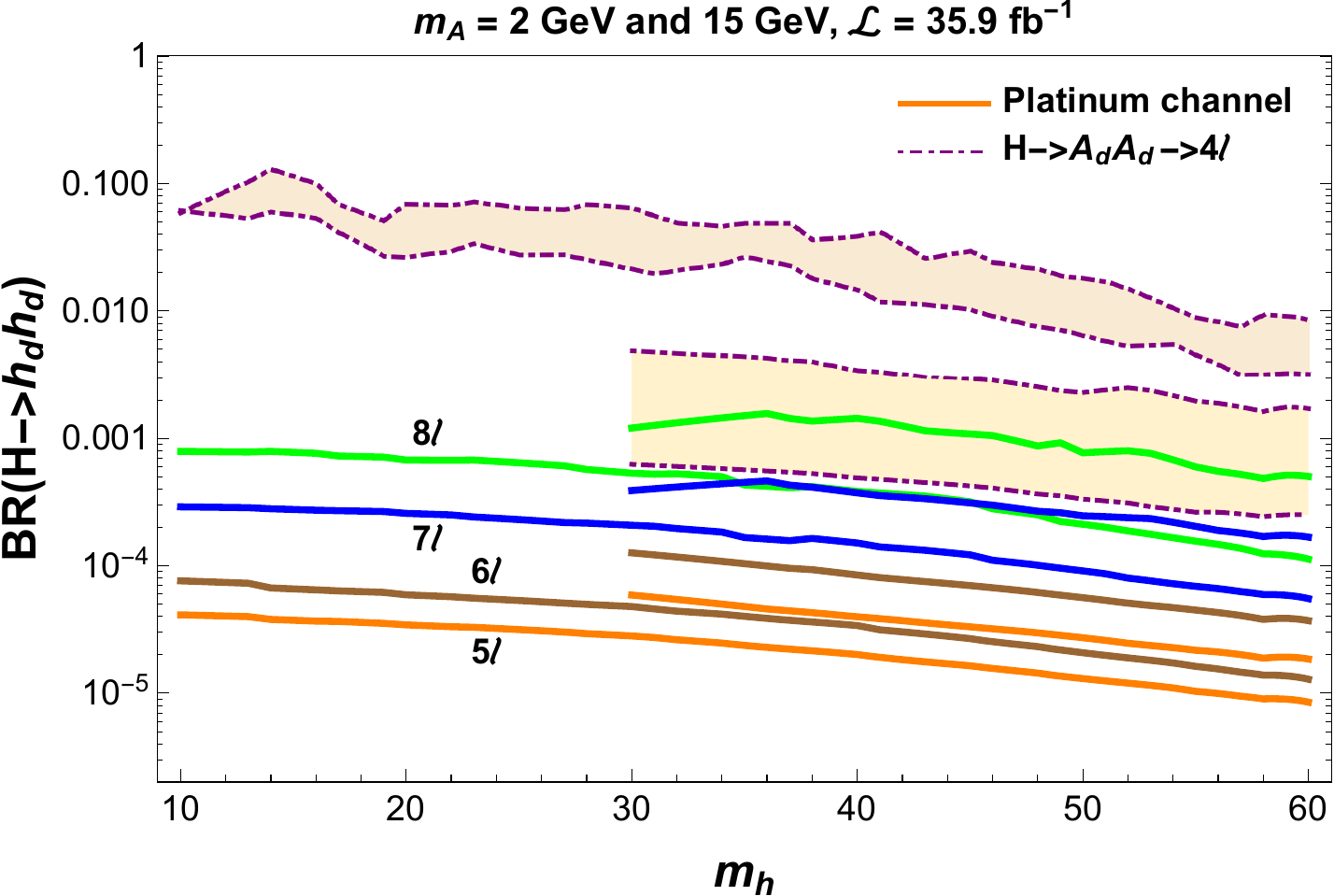}
\hfill
\caption{Current and projected limits on BR($H\rightarrow h_dh_d$) as a function of the mass of the hidden scalar. Here we show only limits and projections directly on this production mode, and the colours are the same as in \fref{Haa}. The lines that go all the way across are for $m_{A_d} = 2$ GeV, while those that stop at 30 GeV are for  $m_{A_d} = 15$ GeV. Projections are for 35.9 fb$^{-1}$.}
\label{fig:Hhh}
\end{figure*}
%

\section{Strategies and Projections}
\label{sec:projections}

We now comment on potential for improvement with a dedicated analysis. We focus on multilepton final states beyond four leptons because $5^+$ lepton final states should have low backgrounds.  It is beneficial to use
multilepton triggers with low $p_T$ thresholds. Currently, the three-lepton triggers seem like a good candidate, given the low $p_T$ requirements on the leptons. For ATLAS, these are given by~\cite{ATL-DAQ-PUB-2017-001}
\begin{itemize}
  \item three loose $e$'s: $p_T \geq 15,\; 8,\; 8$ GeV at L1 (17, 10, 10 at HLT),
  \item three $\mu$'s: $p_T > 6~\rm{GeV}$ ($3\times6$ at HLT).
\end{itemize}
For CMS, a multilepton analysis~\cite{CMS:2014xja} used 
\begin{itemize}
  \item three $e$'s: $p_T \geq 15,\; 8,\; 5$ GeV.
\end{itemize}
While these analyses were performed at 8 TeV, the trigger thresholds did not increase significantly in the 13 TeV run
\cite{Khachatryan:2014qwa,Sirunyan:2017lae}, so we use these thresholds for our estimated projections. For the leptons in addition to those required to pass the trigger, we require $p_T(\mu) > 2$ GeV~\cite{CMS-DP-2017-029} and $p_T(e) > 5$ GeV
~\cite{Khachatryan:2015hwa}. For all electrons (muons), we require $|\eta| < 2.5 \;(2.4)$.  We also require that the leptons are isolated using the $p_T$ dependant isolation criteria from~\cite{Sirunyan:2017lae}. In order to reduce the background, we further require:
\begin{itemize}
\item $m_\text{all} < 130$ GeV
\item $m_\text{OSSF}\not\in [0,1.1]\cup[2.7,3.8]\cup[9.1,11.1]$ GeV
\end{itemize}
where $m_\text{all}$ is the invariant mass of all reconstructed isolated leptons, and its required to be near or below the Higgs mass. This reduces background from processes with top quarks such as $\bar{t}tZ$ to below attobarn (ab) cross sections.

$m_\text{OSSF}$ is the invariant mass \textit{any} pair of leptons with the opposite signs and the same flavor, and it is required to not be near a QCD resonance which can decay to dieleptons and also to not be too low. Given these cuts, the leading background is multi-boson production. We simulate in {\sc MadGraph5\_aMC@NLO} $pp \rightarrow VV \rightarrow 4\ell$ where $V=Z,\gamma$. We then shower and hadronize the events using {\sc Pythia8.2} including QED radiation. This procedure yields a cross section for producing 5 (6) leptons to be 18 (0.7) ab. Backgrounds with fake electrons can be estimated from the jet faking electron rate~\cite{Khachatryan:2015hwa} times the rate of events with 4 leptons, and should be smaller than real $VV$ background. Fake muons are smaller still. 

We can then place a projected limit for ${\cal L} = 35.9$ fb$^{-1}$ assuming there will 0.63 (0) expected background events with $\geq n$ leptons with $n=5$ ($n=6,7,8$). Since all channels have small backgrounds, the $\geq$5 lepton channel will have the best projected limit, but we show all four possible values of $n$ to motivate different possible searches. In particular, an excess in the $n = 8$ lepton bin is particularly interesting as it allows to potentially fully reconstruct the Higgs invariant mass. We show the projected limit BR($H\rightarrow h_dh_d$) in \fref{Hhh}. For low mass $A_d$, a dedicated search along these lines would exceed current limits by about three orders of magnitude, while for moderate mass $A_d$ by a factor of a few. 

This projected limit assumes a luminosity of 35.9 fb$^{-1}$ at 13 TeV, the same amount of data used in~\cite{Sirunyan:2017lae}, and much less than the total amount of data presently collected. We see that even with this modest integrated luminosity, branching ratios of $H\rightarrow h_dh_d$ as low as $\mathcal{O}(10^{-5})$ can be explored. At higher integrated luminosity, rare background processes will become more important, but we can still expect significant improvements with more data.

We also show the projected bound on BR($H\rightarrow A_dA_d$) in \fref{Haa} using Eqs.~\eqref{eq:Haa} and~\eqref{eq:Hhh}. These bounds are comparable to the recent 13 TeV ATLAS result~\cite{Aaboud:2018fvk}, but we stress that this comparison only applies the minimal model, and relative decay rates of the Higgs to vectors vs.~scalars will be modified in non-minimal models.

\section{Non-Minimal Models}
\label{sec:nonmin}

In the simple model presented in Sec.~\ref{sec:model}, the Higgs decay to hidden two vectors and to two hidden scalars have correlated rates as shown in \eref{Haa} and \eref{Hhh}. Nature need not realize such a simple model, and, as seen in \fref{Haa}, the strongest current constraint on this simple model in much of the parameter range is from the Higgs decay to two vectors which then go to four leptons. Therefore, we here present a simple extension where the decay to two hidden photons can be parametrically smaller than the decay to two hidden scalars. 

Consider a model with a $U(1)$ as above, but with two hidden scalars that have unit charge under the $U(1)$, $h_1$ and $h_2$, that are still neutral under all SM gauge groups. As in the 2HDM, the Higgs potential now has many more parameters and there are potentially multiple new and interesting processes that can arise. Here we will do a simplified analysis assuming that the mixing between $h_1$ and $h_2$ is small, and 
\begin{eqnarray}
m_{h_2} >  m_H/2 > m_{h_1}\, , \;\;\;\; v_2 \gg v_1\, , \;\;\;\; \lambda_2 \ll \lambda_1 \ll 1\, , \;\;\;\;
\label{eq:2hparam}
\end{eqnarray}
where $m_H = 125$ GeV is the mass SM-like Higgs, $v_i$ is the vev of the $i$th hidden Higgs, and $\lambda_i$ is the scalar portal coupling of the $i$th hidden Higgs to the SM-like Higgs.  

In this regime, the SM Higgs decay to two hidden vectors through mixing with $h_1$ ($h_2$) is suppressed by the small parameter $v_1$ ($\lambda_2$). The decay to two $h_2$'s is forbidden by kinematics, while the decay to two $h_1$'s goes like $\lambda_1 v$, where $v\simeq 246$ GeV is the SM Higgs vev. The decay of the $h_1$ to two hidden vectors is suppressed by the small parameter $v_1$, but if there is nothing else that the $h_1$ can decay to, this will not be a suppression to the rate of the SM Higgs decay to four hidden vectors and the platinum channel decay of \eref{decay}.

In this more complicated scenario, the strongest bound will be from the CMS multi-lepton search, and from \fref{Hhh} we see that the branching ratio of $H\rightarrow h_dh_d$ can be as large as 10\% for low $m_{h_d}$ and $m_{A_d}$. In that scenario, the proposed search of this work would improve bounds even more than in the minimal model.  We here stress that the parameter region of \eref{2hparam} is simply an existence proof of a relatively simple model where the decay to two vectors can be suppressed while the decay to four vectors can be large. We leave a full study of the more complicated models and other extensions to future work.

\section{Summary and Conclusions}
\label{sec:conclusion}

Hidden massive photons have recently generated significant interest in the community and spurred significant experimental progress~\cite{Battaglieri:2017aum}. If such a photon gets mass from a Higgs mechanism, then one naturally expects a Higgs portal coupling between the hidden Higgs and the SM Higgs. In such a scenario, if the dark vector and scalar are near the weak scale, then the SM like Higgs could easily have a decay
\begin{equation}
H\rightarrow h_d h_d \rightarrow 4A_d\;.
\end{equation}
The $A_d$ could in turn decay to a pair of leptons, allowing for Higgs decays with final state with large numbers of leptons. Such a signature would be spectacular at the LHC and laregly background free. 

While there are some searches for many leptons, there is no dedicated search for Higgs decay in this channel, and current searches are relatively weak. A dedicated search requiring at least five leptons can significantly increase the reach for such a scenario, with \fref{Hhh} showing a reach with a branching ratio of the SM Higgs to two hidden scalars as low as $10^{-5}$ using the 36 fb$^{-1}$ of data that has already been analyzed. Significant improvements are expected with higher luminosity, especially for $n=7,8$ leptons where the backgrounds should be negligible even at the high-luminosity LHC.

Finally, we note that a genuine experimental study is needed to make precise predictions on the reach. The sensitivity depends on lepton thresholds, the lower the better. At very low thresholds, however, experimental issues such as fakes become significantly more difficult, so we note that with the low thresholds used in this study, the uncertainties on our projections will be relatively large. Yet given the significant gains possible with a dedicated search and the simplicity of the model presented here, we believe that such a search may be well worth the effort.

~\\
\noindent
{\bf Acknowledgments:}~We thank Nikita Blinov, Will Buttinger, David Curtin, Christopher Hayes, Roberto Vega-Morales, and Rachel Yohay for comments on the draft. We are also grateful to the referee from PRL who spurred a more thorough background study. EI is supported by the United States Department of Energy under Grant Contract desc0012704. DS is supported in part by the Natural Sciences and Engineering Research Council of Canada (NSERC) and is grateful for the hospitality of Brookhaven National Laboratory where this work began.


\bibliographystyle{apsrev}
\bibliography{references}

\end{document}